\documentclass[lettersize,journal]{IEEEtran}
\usepackage{amsmath,amsfonts}
\usepackage{algorithmic}
\usepackage{algorithm}
\usepackage{algorithmic}
\usepackage{array}
\usepackage[caption=false,font=normalsize,labelfont=sf,textfont=sf]{subfig}
\usepackage{textcomp}
\usepackage{stfloats}
\usepackage{url}
\usepackage{verbatim}
\usepackage{graphicx}
\usepackage{cite}

\usepackage{amsmath}
\usepackage{bm}
\usepackage{dsfont}
\usepackage{hyperref}

\begin{document}

\title{Deploying Graph Neural Networks in Wireless Networks: A Link Stability Viewpoint}

\author{
Jun Li, Weiwei Zhang, Kang Wei, Guangji Chen, Long Shi, Wen Chen
\thanks{

Jun Li, Weiwei Zhang, Guangji Chen, and Long Shi are with the School of Electronic and Optical Engineering, Nanjing University of Science and Technology, Nanjing 210094, China(e-mail: \{jun.li, wwzhang, guangjichen\}@njust.edu.cn and slong1007@gmail.com)

Kang Wei is with the Department of Computing, The Hong Kong Polytechnic University, Hong Kong 100872, China (e-mail: kangwei@polyu.edu.hk)

Wen Chen is with the Department of Electronic Engineering, Shanghai Jiao Tong University, Shanghai 200240, China (e-mail: wenchen@sjtu.edu.cn)

}
}

\markboth{IEEE WIRELESS COMMUNICATION LETTERS}%
{Shell \MakeLowercase{\textit{et al.}}: A Sample Article Using IEEEtran.cls for IEEE Journals}


\maketitle

\begin{abstract}
As an emerging artificial intelligence technology, graph neural networks (GNNs) have exhibited promising performance across a wide range of graph-related applications.
However, information exchanges among neighbor nodes in GNN pose new challenges in the resource-constrained scenario, especially in wireless systems. 
In practical wireless systems, the communication links among nodes are usually unreliable due to wireless fading and receiver noise, consequently resulting in performance degradation of GNNs. 
To improve the learning performance of GNNs, we aim to maximize the number of long-term average (LTA) communication links by the optimized power control under energy consumption constraints. 
Using the Lyapunov optimization method, we first transform the intractable long-term problem into a deterministic problem in each time slot by converting the long-term energy constraints into the objective function. 
In spite of this non-convex combinatorial optimization problem, we address this problem via equivalently solving a sequence of convex feasibility problems together with a greedy-based solver. 
Simulation results demonstrate the superiority of our proposed scheme over the baselines.
\end{abstract}


\section{Introduction}
\IEEEPARstart{W}{ith} 
the explosion of intelligent devices and improved quality of service (QoS) demands, 
the integration of artificial intelligence (AI) techniques becomes increasingly indispensable for sixth-generation (6G) wireless communication systems\cite{Li2022, Wei2023}. 
As an emerging AI technology, graph neural networks (GNNs), designed for graph-structured data, have demonstrated superior performance across various graph-related applications\cite{wu2020comprehensive}. 
GNNs have exhibited superiority in learning the topological characteristics of graphs and strong generalization capabilities that can adapt to different graph structures and task requirements. 

In wireless GNN systems, the information exchanges among neighbor nodes are achieved over wireless channels during the training process of GNNs. 
However, wireless transmissions over fading and noisy channels are inherently unreliable, which may result in transmission outage (TO) of the desired communication link\cite{park2009outage}. 
Furthermore, TO can deteriorate the information exchanges during the training process of GNNs, and inevitably impose adverse impacts on the performance of GNNs.
Besides, for resource-limited users in the system, the limitation of energy is a major issue restricting the efficiency of training tasks. 
Despite the potential benefits of GNNs, resource limitations faced by users and fading as well as noisy channels pose significant challenges to the performance of GNNs, which becomes a crucial bottleneck of wireless GNN systems.

To implement GNNs in wireless systems, the work in ~\cite{Gao2023} integrated the communication model into a GNN architecture, improving robustness to channel impairments by modifying the graph convolutional operation to consider channel fading and noise during feature aggregation over random communication graphs.
The work~\cite{lee2021decentralized} studied robust decentralized inference with GNNs in wireless communication systems and analyzed the performance of the decentralized GNN binary classifier in both uncoded and coded systems.
Then, this work developed two retransmission mechanisms based on the analysis to enhance the robustness of GNN classifier. 
Deploying GNNs in wireless networks, devices are modeled as users exchanging information through links, and the number of effective links is a basic guarantee of GNN learning performance. 
However, the above works failed to take into account the limited energy resources, fading, channel noise, as well as co-channel interference among nodes. 
The presence of these factors contributes to the unreliability of links, and there is no comprehensive framework available for characterizing these factors.

Driven by the aforementioned issues, we aim to maximize the number of activated communication links in the training process of GNNs under the long-term average (LTA) energy constraints. 
To this end, we formulate a stochastic optimization problem to enhance the performance of GNNs under limited energy supply. 
Specifically, we propose to effectively determine the transmit power of nodes to maximize the number of activated links subject to constraints on the peak transmit power and the LTA energy supply.
Then we leverage the Lyapunov optimization method to convert the long-term stochastic optimization problem into a deterministic optimization problem in each time slot.
Although this deterministic combinatorial optimization problem is non-convex and obtaining its optimal solution is challenging in general, we obtain the high-quality solution of this problem via equivalently solving a sequence of convex feasibility problems together with a greedy-based solver. 
Simulation results demonstrate that the proposed algorithm can guarantee the stability of the virtual queue and significantly increase the number of activated links. 
Moreover, we show that our proposed algorithm achieves superior learning accuracy over the baseline strategies.


\section{System model and problem formulation}
Let $\mathcal{G}=\left( \mathcal{V}, \mathcal{E}\right)$ denote a GNN input graph, where $\mathcal{V}$ and $\mathcal{E} \subseteq \mathcal{V} \times \mathcal{V}$ are the set of vertices and edges, respectively. 
The total number of nodes is denoted by $n$. 
We define the node feature matrix as $\mathbf{X}\in \mathbb{R}^{n\times d}$ where each row $x_v\in \mathbb{R}^{d}$ is a $d$-dimension feature of node $v$. 
We define the neighbor set of node $v$ as $\mathcal{N}(v)$, and the degree of node $v$ is the number of neighbors (i.e., $\mathrm{deg}(v)=\vert \mathcal{N}(v) \vert$). 
The topological information of graph $\mathcal{G}$ is described by the adjacency matrix $\bm{A}\in \mathbb{R}^{n\times n}$ where $\bm{A}_{uv}=1$ if $(u,v)\in \mathcal{E}$ and $\bm{A}_{uv}=0$ otherwise. 
Denote the training round index set by $\mathcal{T} = \{1,...,t,...\}$. 
To enable local computation, we assumed that each node possesses sufficient storage and computational resources\cite{sufficientResource}. 
Additionally, nodes can exchange information with neighbors if they can communicate with each other over wireless channels successfully.

\subsection{Wireless Communication Model}
We adopt the time division scheme for the wireless network which partitions each time slot into a set of orthogonal sub-slots, allowing multiple transmission node pairs to transmit within the same slot without interference. 
Specifically, we divide the topology $\bm{A}$ into multiple sub-topologies $\bm{A}'\in \mathbb{R}^{n\times n}$, each corresponding to the topology in a specific sub-slot. 
Furthermore, each node in the sub-topology can only receive or transmit information, that is, $\sum_{u\in \mathcal{V}} A_{uv}' \leq 1$ and $\sum_{v\in \mathcal{V}} A_{uv}' \leq 1$. 
Assuming an orthogonal access scheme where each user can allocate one or more specific sub-slots within a slot to transmit their data, and each user cannot transmit or receive information simultaneously in one sub-slot.
This division of time into sub-slots ensures that users can transmit messages without colliding with each other, thereby enabling efficient use of the available bandwidth.
The wireless channel is assumed to be stationary and invariant within a time slot but varies over different slots~\cite{Chen2023}.

\subsection{Transmission Outage}
In this subsection, we consider the impact of TO on the performance of GNN. 
In wireless communications, TO may occur caused by the attenuation of signal strength\cite{park2009outage}. 

We denote $h_{uv}$ as the channel coefficient of the link from neighbor node $u$ to node $v$. 
The received signal-to-interference-plus-noise ratio (SINR) of the link from the neighbor node $u\in \mathcal{N}(v)$ to node $v$ is given by 
\begin{equation}\label{2}
    \xi_{uv}=\frac{\vert h_{uv} \vert^2 P_u(t)}{\sum_{k \in \mathcal{K}(t),k\neq u }\vert h_{kv} \vert^2 P_k(t)+\sigma_N^2},
\end{equation}
where $\mathcal{K}(t)$ denotes the set of transmit nodes and $\sigma_N^2$ reprsents the power of the addictive white Gaussian noise. 

Accordingly, the achievable rate $C_{uv}(t)$ over the $u$-$v$ link is given by
\begin{equation}\label{3}
C_{uv}(t) = B\log_2(1+\xi_{uv}(t)),
\end{equation}
where $B$ represents the channel bandwidth.

We define the minimum data rate of the communication link from the neighbor node $u\in \mathcal{N}(v)$ to node $v$ as 
\begin{equation}\label{4}
    R_{uv}(t)=D_u(t)/\tau^{\text{max}},
\end{equation}
where $D_u(t)$ denotes the number of bits that node $u$ requires to transmit node feature to neighbor nodes, 
and $\tau^{\text{max}}$ represents the maximum tolerable transmission delay of communication links in the system.

It is known that the TO occurs if the defined minimum transmission rate $R_{uv}(t)\textgreater C_{uv}(t)$, since the receiver cannot accurately decode the information. 
We use a communication indicator vector $\mathds{1}_{uv}(t)$
to denote whether the information is transmitted from node $u$ to neighbor node $v$ successfully.               
If $R_{uv}(t) \leq C_{uv}(t)$, then the transmission is successful, i.e., $\mathds{1}_{uv}(t)=1$. 
Otherwise, TO occurs and $\mathds{1}_{uv}(t)=0$.

\begin{figure}[!t]
\centering
\includegraphics[width=3.5in]{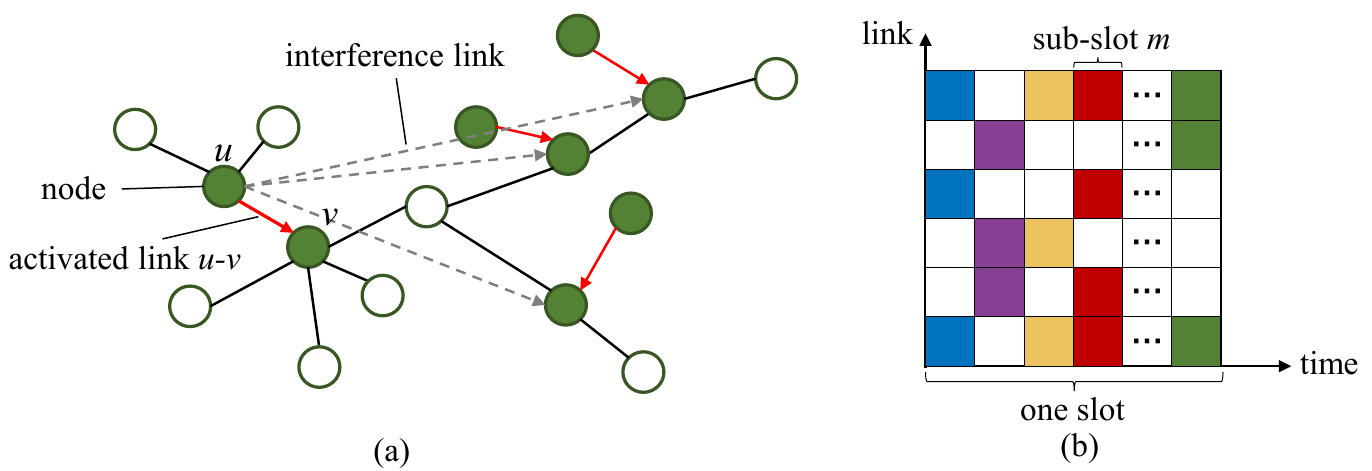}
\caption{Wireless graph neural network system. (a) Node transmission in the system. (b) Orthogonal access in the time division system, where a slot is divided into sub-slots, and each node can only receive or transmit information in each sub-slot.}
\label{system}
\end{figure}

\subsection{Decentralized Training Process of GNN with TO over Wireless Channels}
In this subsection, we deploy GNNs in the wireless system. 
In wireless GNN systems, the information exchanges among neighbor nodes are achieved over wireless channels during the training process of GNNs. 
Without loss of generality, we adopt the representative GNNs, i.e., graph convolutional network (GCN), as the GNN model in the subsequent discussion, the operations of the $k$-th layer of GCN are given by 
\begin{equation}
\begin{aligned}
f_{\text{AGG}}^{(k)} \Big( f_{\text{M}}^{(k)}\big(h_{u}^{(k)} \vert u \in \mathcal{N}(v)\big) \Big) &= \sum_{u\in \mathcal{N}(v)} \frac{h_{u}^{(k)}}{\sqrt{\mathrm{deg}(v)}\sqrt{\mathrm{deg}(u)}},\notag \\
f_{\text{UP}}\big(m_{\mathcal{N}(v)}^{(k)}, h_{v}^{(k)}\big) &=\sigma(\mathbf{W}^{(k)}m_{\mathcal{N}(v)}^{(k)}),
\end{aligned}
\end{equation}
where the aggregation is normalized by degree from transmit and receive nodes (i.e., $\mathrm{deg}(\cdot)$) and $\mathbf{W}^{(k)}\in \mathbb{R}^{d\times d}$ is the learnable weight of the linear transformation layer.

For the $K$-layers GNN, the training process of the GNN in the wireless system is as follows: 
\begin{enumerate}
\item{Node $v$ sends the transmission request to each neighbor node $u\in \mathcal{N}(v)$.}
\item{Considering the potential for TO during transmission, neighbor node $u$ transmits the message $m_{uv}^{(k)}(t) = \mathds{1}_{uv}(t)(f_M^{(k)}\big(h_{u}^{(k)}(t)\big))$ to node $v$ over wireless channels.}
\item{Node $v$ sums the collected neighbor messages $m_{uv}^{(k)}(t)$.}
\item{Node $v$ imposes the update function to get the node embedding $h_{v}^{(k+1)}(t)$.}
\end{enumerate}

At this point, we obtain the final node embedding that can be used for the downstream tasks.

\subsection{Problem Formulation}

During the training process of GNNs, there are two main considerations.  
First, the main idea of GNN is to update the representation of each node by aggregating information from its neighboring nodes. 
Hence, the number of links ensuring successful information exchange is crucial to the final performance of GNNs. 
Second, deploying GNNs in wireless networks is usually affected by 
limited energy resources, channel noise, and co-frequency interference, which easily lead to TO during node information exchange. 
Thus, we aim to maximize the number of activated communication links by effectively determining the transmit power of all nodes under LTA energy constraints to mitigate the occurrence of TO.

We define the energy consumption in the $t$-th time slot of transmitting information from node $u$ to node $v$ as $E_u(t) = P_u(t) \tau_{uv}$, where $\tau_{uv}$ denotes the transmission latency from node $u$ to node $v$. 
As devices are generally energy-limited, we should guarantee the LTA energy consumption below the LTA maximum energy supply, i.e., $\frac{1}{T}\sum\limits_{t\in \mathcal{T}} E_u(t) \leq \overline{E}_u^{\text{max}}$.
The total activated communication links of the system is given by
\begin{equation}\label{5}
    N(t)=\sum_{u=1}^{n} \sum_{v=1,v\neq u}^{n} \mathds{1}_{uv}(t) A_{uv}.
\end{equation}
Then, we aim to maximize the average number of activated communication links in GNNs with the LTA energy consumption constraints by effectively optimizing the transmit power of each node. 
Let $\bm{P}(t)=\left\{P_u(t),\forall u\in \mathcal{V}\right\}$ denote the set of transmit power in the $t$-th slot. 
The corresponding stochastic optimization problem is formulated as
\begin{equation}\label{P1:6}
\begin{aligned}
&\textbf{P1:}~\max\limits_{\bm{P}(t)} \overline{N}= \lim\limits_{T\to \infty}\frac{1}{T}\sum\limits_{t\in \mathcal{T}} N(t)\\
\text{s.t.}\quad&\textbf{C1:}~\mathds{1}_{uv}(t)\in \{0,1\},\forall u,v \in \mathcal{V},\forall t \in \mathcal{T},\\
&\textbf{C2:}~0 \leq P_u(t) \leq P_u^{\text{max}} ,\forall u \in \mathcal{V},\forall t \in \mathcal{T},\\
&\textbf{C3:}~0 \leq \frac{1}{T} \sum\limits_{t\in \mathcal{T}} E_u(t) \leq \overline{E}_u^{\text{max}}, \forall u\in\mathcal{V}, \forall t \in \mathcal{T}.\\
\end{aligned} 
\end{equation}

Note that \textbf{P1} is a long-term stochastic problem with a LTA constraint \textbf{C3} and involving binary integer variables in object function, which makes it intractable and difficult to address.

\section{Link maximization over wireless channels under energy constraints}

\subsection{Lyapunov Optimization for Link Maximization}
In this subsection, we leverage the \textbf{Lyapunov optimization framework} to solve the stochastic optimization problem \textbf{P1}. 
Towards this end, we convert the long-term stochastic optimization problem \textbf{P1} into a series of deterministic optimization problems for each training round. 

To solve \textbf{P1}, we first transform the LTA constraint into the queue stability constraint. Specifically, we introduce the virtual queues as follows: 
\begin{equation}\label{7}
    Z_u(t) = \text{max} \big\{ Z_u(t-1) + E_u(t-1) - \overline{E}_u^{\text{max}}, 0\big\} .
\end{equation}

According to \cite{stablequeue2}, we know that a discrete time queue $Q_u(t)$ is mean rate stable if $\lim_{T\to \infty}\frac{1}{T} \sum\limits_{t\in \mathcal{T}}\mathbb{E}\left\{\vert Q_u(t)\vert \right\}$. 
Thus, we can rewrite \textbf{P1} as
\begin{equation}\label{8}
    \textbf{P2:}\quad\max_{\bm{P}(t)}\overline{N},\quad \text{s.t.}\ \textbf{C1, C2}, \tilde{\textbf{C}}\textbf{3}:
    \lim_{t\to \infty}\frac{\mathbb{E}\left\{\vert Z_u(t)\vert \right\}}{t}.
\end{equation}

Let $\bm{Z}(t)=\left\{Z_u(t),\forall u\in \mathcal{V}\right\}$ denote the set of $Z_u(t)$ in the $t$-th slot. 
The Lyapunov function reveals the congestion state of virtue queues. 
We introduce the Lyapunov function $L\big(\bm{Z}(t)\big)$ and the conditional Lyapunov drift $\Delta L\big(\bm{Z}(t)\big)$ as follows:
\begin{equation}\label{9}
\begin{aligned}
    L\big(\bm{Z}(t)\big) &=\frac{1}{2} \sum_{u\in \mathcal{V}}\big(Z_u(t)\big)^2,\\
    \Delta L\big(\bm{Z}(t)\big) &= \frac{1}{2} \mathbb{E}\Big[L\big(\bm{Z}(t+1)\big)-L\big(\bm{Z}(t)\big)\vert \bm{Z}(t)\Big]. \\   
\end{aligned}
\end{equation}

Substituting (\ref{7}) into (\ref{9}), $\Delta L\big(\bm{Z}(t)\big)$ is upper bound by
\begin{equation}\label{10}
    \Delta L\big(\bm{Z}(t)\big) \leq \Theta + \sum_{u\in \mathcal{V}}\mathbb{E}\Big[\bm{Z}(t) \big(E_u(t)- \overline{E}_u^{\text{max}}\big)\vert \bm{Z}(t)\Big],
\end{equation}
where $\Theta = \frac{1}{2}\big((P_u(t)\tau^{\text{max}})^2 + (\overline{E}_u^{\text{max}})^2\big)$. 

It can be observed that minimizing the upper bound in (9) is equivalent to minimizing the term $\mathbb{E}\Big[\sum_{u\in \mathcal{V}}\bm{Z}(t)E_u(t)\Big]$. 
To this end, we derive the Lyapunov drift-plus-penalty ratio function as follows: 
\begin{equation}\label{11}
    \Delta_V(t)= \mathbb{E}\Big[-VN(t)+\sum_{u\in \mathcal{V}} Z_u(t)E_u(t)\vert \bm{Z}(t) \Big],
\end{equation}
where $V\geq0$ is a parameter to balance the energy consumption and the number of links maximization.

In this letter, we apply the opportunistic expectation minimization method \cite{conditionE} in the number of links maximization problem. 
Therefore, we can make the link maximization and power control decision according to the current environment to minimize 
\begin{equation}\label{12}
    \Delta_V(t)= -VN(t)+\sum_{u\in \mathcal{V}} Z_u(t)E_u(t).    
\end{equation}

We transform the problem of maximizing the LTA number of communication links into minimizing the Lyapunov drift-plus-penalty ratio function $\Delta_V(t)$ in each slot.
Therefore, the long-term optimization problem \textbf{P2} can be rewritten as 
\begin{equation}\label{13}
    \textbf{P3:}\quad\min_{\bm{P}(t)}\Delta_V(t),\quad \text{s.t.} \ \textbf{C1, C2} .
\end{equation}

At this point, we already convert the complicated stochastic optimization problem \textbf{P1} into the deterministic combinatorial optimization problem \textbf{P3} in each slot.
The objective function of \textbf{P3} combines the accumulated energy consumption minimization and the number of activated links maximization. 
Therefore, there exists a tradeoff between the LTA number of activated links and the LTA energy consumption.

\subsection{Greedy-Based Solution to \textbf{P3}}
In this subsection, we propose a greedy-based link maximization algorithm, as shown in Algorithm~\ref{alg1}, to address the stochastic optimization problem \textbf{P3}. 
As described in Section II.A, we adopt the time division scheme for communication.
Nodes are partitioned into $M$ groups such that each node within the same group does not aggregate or transmit information simultaneously.
Accordingly, we divide a time slot into $M$ time slots, with the $m$-th group of nodes communicating with each other in the $m$-th sub-slot.
Then, we solve the combinatorial optimization problem \textbf{P3} in each sub-slot. 

It can be observed that the object function of \textbf{P3} with binary variables is non-convex and challenging to be solved directly by convex optimization techniques. 
Hence, we propose to determine the optimal power control in various cases that retain a different set of activated links. 
Specifically, we begin by sorting the links in descending order according to the achievable rate $C_{uv}(t)$ in the $m$-th sub-slot and denote them as $\{\bm{L}\}_m$, and then progressively eliminate link with lower $C_{uv}(t)$ as different cases. 
In other words, there are $J = \vert\bm{A}_m \vert$ cases in the $m$-th sub-slot, and we define the set of transmit nodes in the $j$-th case as $\mathcal{K}_j(t)$.

In the following, we show that \textbf{P3} can be optimally solved by equivalently solving a sequence of feasibility problems in each case. 
We take judgement condition of TO in the objective of \textbf{P3}, i.e., $ C_{uv}(t)\geq R_{uv}(t)$, as a new constraint $\textbf{C3}^*$ and define the feasibility problems as
\begin{equation}\label{14}
\begin{aligned}
\text{Find}\quad & \{P_u(t)\}\\
\text{s.t.}\quad &\textbf{C2},\\
&\textbf{C3}^*:B\log_2(1+\xi_{uv}(t))\geq D_u(t)/\tau^{\text{max}}.
\end{aligned}
\end{equation}

Recall that $\textbf{C3}^*$ indicates information can be successfully transmitted over wireless channels. 
We rewrite $\textbf{C3}^*$ as 
\begin{equation}\label{15}
\begin{aligned}
\vert h_{uv} \vert^2 P_u(t)-\sum_{k \in \mathcal{K}_j(t),k\neq u }\vert h_{kv} \vert^2 P_k(t) G&\geq G\sigma_N^2,\\
\end{aligned}
\end{equation}
where $G=2^{D_u/B\tau^{\text{max}}}-1$.

At this point, \textbf{P3} is reformulated as the following convex feasibility problem, which can be solved by convex optimization tools, e.g., CVX. 
\begin{equation}
\begin{aligned}\label{16}
    \text{Find}\quad & \{P_u(t)\}\\
\text{s.t.}\quad &\textbf{C2},\\
&\vert h_{uv} \vert^2 P_u(t)-\sum_{k \in \mathcal{K}_j(t),k\neq u }\vert h_{kv} \vert^2 P_k(t) G&\geq G\sigma_N^2.\\
\end{aligned}
\end{equation}


Ultimately, we select the $\{\bm{L}\}_m^{\text{opt}}$ and $\bm{P}^{\text{opt}}$ with the smallest objective function value in \textbf{P3} in all cases as the optimal result, and take the $\bm{A}_m^{\text{opt}}$ corresponding to $\{\bm{L}\}_m^{\text{opt}}$ as the topology of the GNN training process in the $m$-th sub-slot.
Following these solution steps in each sub-slot, we can determine the optimal sub-topology in each sub-slot. 
At last, we obtain the final communication topology $\bm{A}^{\text{opt}}(t)$ by combining all sub-topologies $\bm{A}_m^{\text{opt}}(t)$ for GNN training. 
\begin{algorithm}[]
\caption{Greedy-based Link Maximization Algorithm}\label{alg1}
\begin{algorithmic}
\setlength{\algorithmicindent}{0cm}
\STATE$\textbf{Require}$: The channel state ${h_u(t)}$, sub-adjacency matrix $\{\bm{A}_m, \forall m\in M\}$ and virtual queue length $\bm{Z}(t)$;
\STATE$\textbf{Ensure}$: The sub-adjacency matrix $\bm{A}_m$
\STATE1: Initialize: $l$ = 0,  $\bm{Z}(t)=0$, $P^{\text{max}}$, $\overline{E}_u^{\text{max}}$
\STATE2: \textbf{while} $t<T$ \textbf{do}
\STATE3: \hspace{0.5cm} \textbf{for} each sub-slot \textbf{do}
\STATE4: \hspace{1cm}Sort all links of $\bm{A}_m$ in descending order according\\
\hspace{1.25cm} to (2) as $\{\bm{L}\}_m$;
\STATE5: \hspace{1cm}\textbf{while} $\{\bm{L}\}_m$ not null \textbf{do}
\STATE6: \hspace{1.5cm}Remove the $l$-th link;
\STATE7: \hspace{1.5cm}\textbf{if} problem (15) is feasible \textbf{then}
\STATE8: \hspace{1.5cm}\hspace{0.5cm} Find $\{P_u(t)\}$ by solving problem (15);
\STATE9: \hspace{1.5cm}\textbf{end if}
\STATE10:\hspace{1.45cm}Set $l$ = $l+1$;
\STATE11:\hspace{1cm}\textbf{end while}
\STATE12:\hspace{1cm}Choose $\bm{A}_m^{\text{opt}}$ and $\bm{P}^{\text{opt}}(t)$ that minimize the 
\STATE\hspace{1.45cm}objective of \textbf{P3} as the optimal solution in the $m$-th\\
\STATE\hspace{1.45cm}sub-slot;
\STATE13:\hspace{1cm}Update $\bm{Z}(t)$ according to (6);
\STATE14: \hspace{0.5cm} \textbf{end for}
\STATE15: \hspace{0.5cm} Obtain $\bm{A}^{\text{opt}}(t)=\sum_{m=1}^{M}\bm{A}_m^{\text{opt}}(t)$;
\STATE16: \hspace{0.5cm} Set $t$ = $t+1$;
\STATE17: \textbf{end while}
\STATE18: \textbf{return} $\bm{P}^{\text{opt}}(t)$, $\bm{A}^{\text{opt}}(t)$ 
\end{algorithmic}
\end{algorithm}



\section{Simulation results}
In this section, we provide the simulation and analytical results of the wireless GNN system to demonstrate the performance of the proposed algorithm. 

We conduct the simulation on the popular graph datasets, Cora~\cite{cora} and Citeseer~\cite{citeseer}, which are benchmark datasets for node classification tasks. 
In these two datasets, nodes correspond to scientific papers and edges to citations. Each paper is represented by a bag-of-words feature vector and labeled with different research fields. 
Cora contains 2708 nodes, 5429 edges, 7 classes, and 1433 features of each node. 
Citeseer consists of 3327 nodes, 4732 edges, 6 classes, and 3703 features per node.
We randomly split the dataset into three potions, 60 $\%$ for training, 20$\%$ for validation, and 20 $\%$ for testing.

The path loss is 30.6+36.7$\log_{10}(d)$, with $d$ in meter, while Rayleigh fading is used for small scale fading. 
Unless otherwise stated, we set $V=0.00001$.
We adopt a 2-layer GCN to conduct node classification tasks on Cora and Citeseer datasets. 
Each of the two layers has a hidden dimension of size 16 and a ReLU activation function.
Other experimental parameters are listed in Table~\ref{tab:table1}.

For comparison purposes, we compare the proposed algorithm with two baseline strategies as follows: (a) all nodes are transmitting information by the maximum transmit power (abbreviated as ``Full Power''), 
(b) The energy consumption of each node in each slot is constrained, meaning that the LTA energy constraints \textbf{C3} in \textbf{P3} is replaced with $\hat{\textbf{C}}3:0 \leq E_u(t) \leq \overline{E}_u^{\text{max}}, \forall u\in\mathcal{V}, \forall t \in \mathcal{T}$ (abbreviated as ``ECPS'').
Additionally, we adopt the same GCN without any optimization strategy as the upper bound of accuracy performance.

\begin{table}[!t]
\caption{simulation parameters\label{tab:table1}}
\centering
\begin{tabular}{c|c||c|c||c|c}
\hline
Parameter & Value & Parameter & Value & Parameter & Value \\
\hline
$B$ & 1 MHz & $P^{\text{max}}$ & 0.5 W &  $d$ & 100 m\\
\hline
$\overline{E}_u^{\text{max}}$ & 1.25 J & $D_u$ & 1 Mbit & $\tau^{\text{max}}$ & 0.5 s \\
\hline
\end{tabular}
\end{table}

\begin{figure}[!t]
\centering
\includegraphics[width=3.5in]{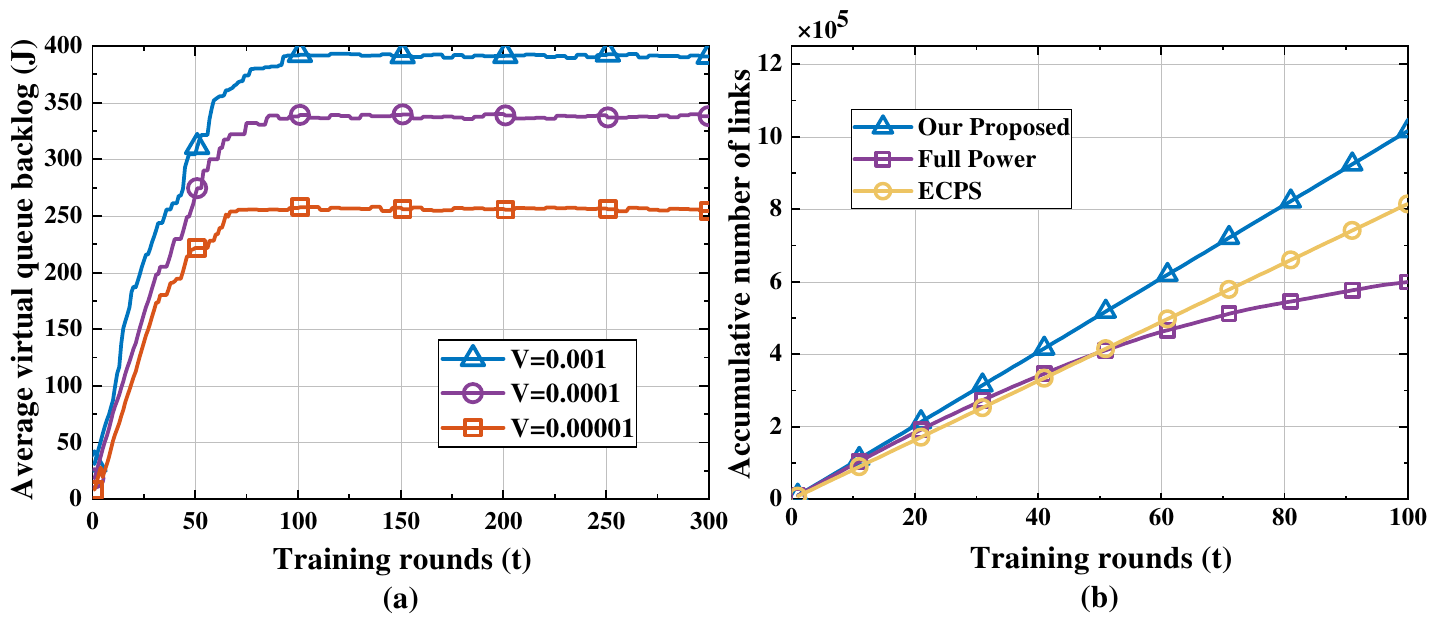}
\caption{Average virtual queue backlogs with different $V$ values and 
the accumulative number of links on Cora, respectively.}
\label{z_link}
\end{figure}

In Fig.~\ref{z_link}(a), 
we compare the proposed algorithm with two baseline strategies in terms of the average virtual queue backlogs. 
Fig.~\ref{z_link}(a) illustrates the average virtual backlogs with different values of $V$= 0.001, 0.0001, and 0.00001, respectively. 
It is observed that the average virtual backlogs increase in the beginning and gradually become stable over time. 
Furthermore, we can observe that the average virtual backlog of nodes can reach convergence faster when $V$= 0.00001.
This is due to the fact that we emphasize more on the minimization of energy consumption and less on the maximization of activated communication links when minimizing 
$\delta_V(t)$ in (11) under a relatively small $V$.

\begin{figure}[!t]
\centering
\includegraphics[width=3.5in]{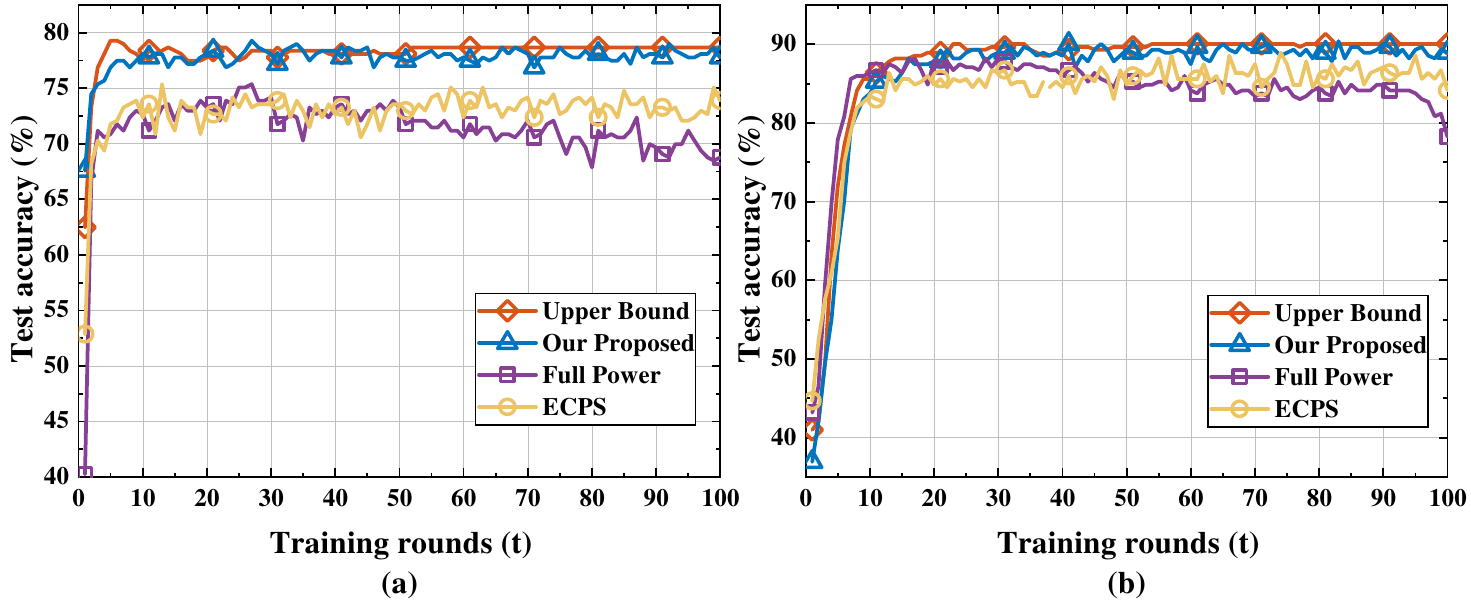}
\caption{The test accuracy of our proposed algorithm and baseline algorithms. (a) and (b) show the test accuracy on Citeseer and Cora, respectively.}
\label{acc}
\end{figure}

Fig.~\ref{z_link}(b) depicts the number of activated communication links of the proposed algorithm and two baselines. 
First, it can be observed in Fig.~\ref{z_link}(b) that the proposed algorithm outperforms baselines and can improve the LTA activated communication links effectively. 
The reason is that the proposed algorithm considers the channel state and energy consumption information rather than the partial information in two baselines. 
The strategy Full Power operates with the maximum transmit power under LTA energy constraints. 
However, it leads to some nodes exhausting the energy budget allocated by the system prematurely, forcing them to cease operation ahead of time. 
Additionally, the strategy ECPS enforcing energy constraints at each time slot leads to the termination of some nodes within some time slot if they consume more energy than the energy supply, resulting in fewer activated links.

As for the evaluation metric of GNN performance, we utilize classification accuracy on the test set to evaluate the generalization capability of the learning model.
Fig.~\ref{acc} compares the test accuracy performance of our proposed algorithm on Citeseer and Cora, respectively. 
It can be observed that the proposed algorithm achieves better performance than the other two baseline strategies. 
This is because our proposed algorithm takes into account power control, channel states, and energy consumption of each node to maximize the number of communication links. 
Besides, we leverage a tradeoff between the maximization of link numbers and the energy consumption of nodes in each slot. 
However, the Full Power and the ECPS strategies only consider power or energy consumption in each slot, which results in worse learning performance than our proposed algorithm.  

\section{Conclusion}
In this letter, we have studied the number of activated links maximization in the process of GNNs training under maximum power constraint and LTA energy constraint. 
Based on the Lyapunov optimization method, we have introduced the virtual queue for each node's energy consumption and converted the long-term stochastic optimization problem into the deterministic combinatorial optimization problem in each slot.
Subsequently, we leveraged the greedy-based algorithm to find the optimal solution to the combinatorial optimization problem. 
The simulation results have unveiled the stability of the virtual queue and the effectiveness of increasing the number of activated links. 
In addition, it has been demonstrated that our proposed algorithm can achieve superior learning performance than two baseline strategies.     
In the future, the development of the synchronization mechanism between nodes on non-orthogonal channels deserves further research.

\bibliographystyle{IEEEtran}
\bibliography{ref2}







\vfill

\end{document}